\documentclass[aps,amsmath]{revtex4}
\usepackage{amssymb}
\usepackage{graphicx}% Include figure files
\usepackage{bm}% bold math
\begin{document}

\title{The Lorentz Extension as Consequence of the Family Symmetry}
\author{ Hai-Jun Wang}
\address{Center for Theoretical Physics and School of Physics, Jilin
University, Changchun 130023, China}

\begin{abstract}
In this paper we postulate an algebraic model to explain how the
symmetry of three lepton species plays its role in the Lorentz
extension. Inspired by the two-to-one mapping between the group
$SL$(2, C) and the Lorentz group, we design a mapping between
$SL$(3, C) group, which displays the family symmetry, and a
generalized Lorentz group. Following the conventional method, we
apply the mapping results to Dirac equation to discuss its
transformation invariance, and it turns out that only when the
vertex matrix $\gamma _\mu$ is extended to the combination $c_1
\gamma _\mu   + c_2 \gamma _5 \gamma _\mu$ can the
Dirac-equation-form be reserved. At the same time we find that the
Lorentz group has to be extended with an additional generator
$\gamma _5$. The generalized vertex matrix is helpful in
understanding the axial-like form of weak interaction and the
neutrino oscillations.
\end{abstract}

\maketitle

\section{Introduction}

One mystery of particle physics is that the families of quarks and
leptons are just three. Many attempts [1-7] have been made to
explain the origin of the families, most of which [1-6] start from
a larger group and generate the spinors (families) via the
decomposition and breakdown of the group. In Ref. [7, 8] a group
$SU(3)_F$ (or alternatively, $S_3$) is assumed to govern the
underlying family symmetry. Moreover it is elevated to be gauge
symmetry as an extension of the unification symmetry, which could
be broken only at or above unification scale $m_U  = 10^{19}$GeV.

It is an empirical fact that the unitary groups are responsible
for the symmetries of fundamental interactions. And the Lorentz
group is another important sort of group which governs the
transformation of spinors in quantum fields. If the group
governing the family symmetry does indeed rely on the extension of
unitary group, then a new kind of interaction has to be
introduced. However, there are no clues of its existence up to
date, meanwhile the Lorentz extension (violation) in
weak-interaction sector is not yet ruled out by experiments. To
accept family symmetry as a physical fact, in this paper we adopt
the group $SL$(3, C), which can be graded by $SU(3)_F$ (the
concept of "grading" will be explained later), as the underlying
symmetry of family and design a mapping to map it to an
\emph{extended Lorentz group}.

Many works [9-12] have been done on the possibility that the
Lorentz symmetry can also be broken at unifying scale. In Ref. [9]
the relationship between neutrino oscillations and Lorentz
violation was constructed. Here we suppose the three subgroups of
$SL$(3,C), which can be graded by three $SU$(2) group
respectively, form three \emph{Lorentz zones}, because every
subgroup can be mapped to Lorentz group. Each sort of neutrinos
belongs to a Lorentz zone. Through this way we present a picture
based on which the neutrinos oscillate between zones. Since there
exists only one Lorentz zone (group) in nature, the three zones
must be compacted into one by some manner. We construct the manner
by a mapping. The mapping produces the degeneration of the zones
as well as the Lorentz violation (Lorentz extension---It turns out
that the Lorentz group is broken only by adding one more
generator), in coincidence with the hypothesis in Ref. [9]---the
Lorentz violation produces the neutrinos' oscillation.

Since the Standard Model successfully describes most low energy
particle physics, it is reasonably assumed that the above
symmetries' breaking must have some effects at low energies in
terms of effective quantum field theories encompassing Standard
Model. The effects are expected to be detectable in some existing
experiments. In this paper we try to find a way relating the
family symmetry to the extended Lorentz symmetry \emph{in the low
energy limit}. To do so we directly adopt \emph{the SL(3,C)
symmetry} as the family symmetry and at the same time propose a
mapping to associate it with the extended Lorentz group. The
mapping is analogous to that between the group $SL$(2, C) and the
Lorentz group. We divide the $SL$(3,C) group into three
correlative $SL$(2,C) subgroups by their grading manners and find
it able to be mapped into three Lorentz groups. While making one
$SL$(2, C) group as \emph{mapping basis} (the concept
"\emph{mapping basis}" will be elucidated later), it is found that
only one more generator appears to the original Lorentz group. In
such a way the Lorentz group is extended.

The remainder of the paper is arranged as follows: in the next
section, we briefly review the mapping between the group $SL$(2,
C) and Lorentz group. In the third section, we propose a similar
mapping between the group $SL$(3, C) and extended Lorentz group.
In the fourth section we apply the mapping results to Dirac
Equation. Finally the concluding remarks are presented.

\section{Brief review of the relationship between the group \emph{SL} (2, C) and the Lorentz group}

Each element of group $SL$(2, C) has the form

\begin{equation}
 g = \left( {\begin{array}{*{20}c}
   a & b \\
   c & d  \\
\end{array}} \right),
\end{equation}
where $a$, $b$, $c$, $d$ are complex numbers and $\det g = ad - bc
= 1$, which means $Re(\det g )=1 $ and $Im(\det g )=0 $ and thus
leaves the degree of freedom of $g$ to be 6. Equivalently Eq. (1)
can be expressed in the form
\begin{equation}
 g = g_\mu  \sigma ^\mu,
\end{equation}
where $g_\mu$($\mu = 0,1,2,3$) are complex numbers, and
\begin{equation}
 \sigma ^0  = \left( {\begin{array}{*{20}c}
   1 & 0  \\
   0 & 1  \\
\end{array}} \right), \sigma ^1  = \left( {\begin{array}{*{20}c}
   0 & 1  \\
   1 & 0  \\
\end{array}} \right),\sigma ^2  = \left( {\begin{array}{*{20}c}
   0 & { - i}  \\
   i & 0  \\
\end{array}} \right),\sigma ^3  = \left( {\begin{array}{*{20}c}
   1 & 0  \\
   0 & { - 1}  \\
\end{array}} \right).
\end{equation}
We introduce a matrix form $Q$ related to coordinates of
space-time $x_\mu   = (x_0 ,\vec x)$ as follows
\begin{equation}
Q = x_\mu  \sigma ^\mu,
\end{equation}
obviously it gives
\begin{equation}
\det Q = x_0^2  - x_1^2  - x_2^2  - x_3^2.
\end{equation}

We further define the transformation of $Q$ like
\begin{equation}
Q' = gQg^\dag,
\end{equation}
where $Q' = x'_\mu  \sigma ^\mu$, and $x_\mu$ transforms according
to special relativity,
\begin{equation}
x'_\mu   = \Lambda _{\;\,\mu} ^\alpha  x_\alpha.
\end{equation}
From Eqs. (5), (7) we conclude $\det Q$($=\det Q'$) an invariant
quantity. From Eqs. (4), (6), the transformation of $x_\mu$ yields
\begin{equation}
\begin{array}{l}
 x'^\alpha   = \delta _{\;\,\beta} ^\alpha  x'^\beta   = \frac{1}{2}Tr(\sigma ^\alpha  \sigma ^\beta  )x'_\beta   = \frac{1}{2}Tr(\sigma ^\alpha  Q') \\
  = \frac{1}{2}Tr(\sigma ^\alpha  gQg^\dag  ) = \frac{1}{2}Tr(\sigma ^\alpha  g\sigma ^\beta  g^\dag )x_\beta   \\
 \end{array}.
\end{equation}
Then comparing Eq. (8) with the transformation Eq. (7), we can
obtain the mapping between the Lorentz group and the elements  of
\emph{SL}(2, C),
\begin{equation}
\Lambda ^{\alpha \beta }  = \frac{1}{2}Tr(\sigma ^\alpha  g\sigma
^\beta  g^\dag),
\end{equation}
for example, $\Lambda ^{00}  = \left| {g_0 } \right|^2  +
\sum\limits_{k = 1}^3 {\left| {g_k } \right|^2 }$. The reverse of
the expression (9) yields [13]
\begin{equation}
g = g_0 \sigma ^0  + \sum\limits_{k = 1}^3 {g_k \sigma _k }  = D^{
- 1} [Tr\sigma ^0  + \sum\limits_{k = 1}^3 {\Lambda _0^k }  +
\Lambda _k^0  - i\varepsilon _\lambda ^{0k\rho } \Lambda _\rho
^\lambda  ]\sigma ^k,
\end{equation}
where $D^2  = 4 - Tr\Lambda ^2  + (Tr\Lambda )^2  - i\varepsilon
_{\rho \tau }^{\mu \lambda } \Lambda _\lambda ^\tau  \Lambda _\mu
^\rho$.

According to Eqs. (9), (10), the following Lorentz transformations
and elements of $SL$(2, C) can be mutually carried out:
\begin{equation}
{ {\begin{array}{*{20}c}
   $elements of SL (2, C)$  $~~~~~~~~~~~~~~~~~$& $Lorentz transformations$  \\
   \pm\left( {\begin{array}{*{20}c}
   {\cosh \frac{\kappa }{2}} & {\sinh \frac{\kappa }{2}}  \\
   {\sinh \frac{\kappa }{2}} & {\cosh \frac{\kappa }{2}}  \\
\end{array}} \right) $~~~~~~~~~~~~~~~~~$& \left( {\begin{array}{*{20}c}
   {\cosh \kappa } & {\sinh \kappa } & 0 & 0  \\
   {\sinh \kappa } & {\cosh \kappa } & 0 & 0  \\
   0 & 0 & 1 & 0  \\
   0 & 0 & 0 & 1  \\
\end{array}} \right)  \\
 \pm \left( {\begin{array}{*{20}c}
   {\cosh \frac{\kappa }{2}} & {i\sinh \frac{\kappa }{2}}  \\
   { - i\sinh \frac{\kappa }{2}} & {\cosh \frac{\kappa }{2}}  \\
\end{array}} \right) $~~~~~~~~~~~~~~~~~$& \left( {\begin{array}{*{20}c}
   {\cosh \kappa } & 0 & {\sinh \kappa } & 0  \\
   0 & 1 & 0 & 0  \\
   {\sinh \kappa } & 0 & {\cosh \kappa } & 0  \\
   0 & 0 & 0 & 1  \\
\end{array}} \right)  \\
   \pm \left( {\begin{array}{*{20}c}
   {\exp [\frac{\kappa }{2}]} & 0  \\
   0 & {\exp [ - \frac{\kappa }{2}]}  \\
\end{array}} \right) $~~~~~~~~~~~~~~~~~$& \left( {\begin{array}{*{20}c}
   {\cosh \kappa } & 0 & 0 & {\sinh \kappa }  \\
   0 & 1 & 0 & 0  \\
   0 & 0 & 1 & 0  \\
   {\sinh \kappa } & 0 & 0 & {\cosh \kappa }  \\
\end{array}} \right)  \\
\pm \left( {\begin{array}{*{20}c}
   {\cos \frac{\kappa }{2}} & {i\sin \frac{\kappa }{2}}  \\
   {i\sin \frac{\kappa }{2}} & {\cos \frac{\kappa }{2}}  \\
\end{array}} \right) $~~~~~~~~~~~~~~~~~$& \left( {\begin{array}{*{20}c}
   1 & 0 & 0 & 0  \\
   0 & 1 & 0 & 0  \\
   0 & 0 & {\cos \kappa } & { - \sin \kappa }  \\
   0 & 0 & {\sin \kappa } & {\cos \kappa }  \\
\end{array}} \right)  \\
   \pm \left( {\begin{array}{*{20}c}
   {\cos \frac{\kappa }{2}} & { - \sin \frac{\kappa }{2}}  \\
   {\sin \frac{\kappa }{2}} & {\cos \frac{\kappa }{2}}  \\
\end{array}} \right)$~~~~~~~~~~~~~~~~~$&
\left( {\begin{array}{*{20}c}
   1 & 0 & 0 & 0  \\
   0 & {\cos \kappa } & 0 & {\sin \kappa }  \\
   0 & 0 & 1 & 0  \\
   0 & { - \sin \kappa } & 0 & {\cos \kappa }  \\
\end{array}} \right)

  \\

 \pm \left( {\begin{array}{*{20}c}
   {\exp [i\frac{\kappa }{2}]} & 0  \\
   0 & {\exp [ - i\frac{\kappa }{2}]}  \\
\end{array}} \right)

 $~~~~~~~~~~~~~~~~~$& \left( {\begin{array}{*{20}c}
   1 & 0 & 0 & 0  \\
   0 & {\cos \kappa } & { - \sin \kappa } & 0  \\
   0 & {\sin \kappa } & {\cos \kappa } & 0  \\
   0 & 0 & 0 & 1  \\
\end{array}} \right) \\

\end{array}} }
\end{equation}
where the first three lines represent the Lorentz boost, and last
three lines represent the Lorentz rotation, and $\kappa$ stands
for the rapidity. Notice that all the elements in the first column
are independent and can be mutually expressed with the form in Eq.
(2). Therefore the two columns can be seen as the \emph{generating
elements} of the $SL$(2, C) and Lorentz group. We can have the
generators for the two groups by performing the derivatives with
respect to $\kappa$ on these matrices, then making $\kappa = 0$.

\section{The relationship between group \emph{SL}(3,C) and the Lorentz group}

A grading $\Xi$ of a Lie algebra $L$ is a decomposition of $L$
into a direct sum of non-zero grading subspaces $L_j  \subseteq
L$, i.e. $\Xi$:$L =  \oplus _{j \in J} L_j$, such that, for each
pair $j,k$ of indices from the index set $J$, there exists $l \in
J$ with the property $[L_j ,L_k ] \in L_l$. In preceding section
we have seen that the group $SL$(2,C) can be graded by $2 \times
2$ Pauli-matrices plus unit matrix $\sigma ^0  =I_{2\times2}$[14].
Now it is known that $SL$(3,C) group has Pauli gradings [14, 15,
16]. And since it is easy to prove that the Pauli grading matrices
and the generators of the group $SU$(3), e.g. Gell-Mann matrices,
can be mutually expressed linearly, we recognize that $SL$(3,C)
can be graded by generators of the group $SU$(3) under Gell-Mann
representation. The Gell-Mann matrices read

\begin{eqnarray}
\lambda _1  =  \left( {\begin{array}{*{20}c}
   0 & 1 & 0  \\
   1 & 0 & 0  \\
   0 & 0 & 0  \\
\end{array}} \right), \lambda _2  = \left( {\begin{array}{*{20}c}
   0 & { - i} & 0  \\
   i & 0 & 0  \\
   0 & 0 & 0  \\
\end{array}} \right),
\lambda _3  = \left( {\begin{array}{*{20}c}
   1 & 0 & 0  \\
   0 & { - 1} & 0  \\
   0 & 0 & 0  \\
\end{array}} \right),\lambda _4  = \left( {\begin{array}{*{20}c}
   0 & 0 & 1  \\
   0 & 0 & 0  \\
   1 & 0 & 0  \\
\end{array}} \right),\\ \nonumber
\lambda _5 = \left( {\begin{array}{*{20}c}
   0 & 0 & { - i}  \\
   0 & 0 & 0  \\
   i & 0 & 0  \\
\end{array}} \right),\lambda _6  = \left( {\begin{array}{*{20}c}
   0 & 0 & 0  \\
   0 & 0 & 1  \\
   0 & 1 & 0  \\
\end{array}} \right), \lambda _7  = \left( {\begin{array}{*{20}c}
   0 & 0 & 0  \\
   0 & 0 & { - i}  \\
   0 & i & 0  \\
\end{array}} \right), \lambda _8  = \left( {\begin{array}{*{20}c}
   1 & 0 & 0  \\
   0 & 1 & 0  \\
   0 & 0 & { - 2}  \\
\end{array}} \right),
\end{eqnarray}
we can divide the SU (3) group into three relevant parts by
grouping the generators as follows

\begin{equation}
\Gamma _1  = \{ \lambda _1 ,\lambda _2 ,\lambda _3 \}, \Gamma _2
= \{ \lambda _4 ,\lambda _5 ,\left( {\begin{array}{*{20}c}
   1 & 0 & 0  \\
   0 & 0 & 0  \\
   0 & 0 & { - 1}  \\
\end{array}} \right)\}, \Gamma _3  = \{ \lambda _6 ,\lambda _7 ,\left(
{\begin{array}{*{20}c}
   0 & 0 & 0  \\
   0 & 1 & 0  \\
   0 & 0 & { - 1}  \\
\end{array}} \right)\},
\end{equation}
where $\left( {\begin{array}{*{20}c}
   1 & 0 & 0  \\
   0 & 0 & 0  \\
   0 & 0 & { - 1}  \\
\end{array}} \right)$ and $\left( {\begin{array}{*{20}c}
   0 & 0 & 0  \\
   0 & 1 & 0  \\
   0 & 0 & { - 1}  \\
\end{array}} \right)$ can be derived by appropriately combining $\lambda _3$ and $\lambda _8$.
It is obvious that $\Gamma _1$, $\Gamma _2$, $\Gamma _3$ are bases
for three $SU$(2) groups satisfying the commutations of Pauli
matrices. Enlightened by the Eqs. (2), (3), adding three matrices
$I_{3 \times 3}^1  = \left( {\begin{array}{*{20}c}
   1 & 0 & 0  \\
   0 & 1 & 0  \\
   0 & 0 & 0  \\
\end{array}} \right)$, $I_{3 \times 3}^2  = \left( {\begin{array}{*{20}c}
   1 & 0 & 0  \\
   0 & 0 & 0  \\
   0 & 0 & 1  \\
\end{array}} \right)$, $I_{3 \times 3}^3  = \left( {\begin{array}{*{20}c}
   0 & 0 & 0  \\
   0 & 1 & 0  \\
   0 & 0 & 1  \\
\end{array}} \right)$[These three matrices can be derived by appropriately
combining $\lambda _3$, $\lambda _8$ and the unit matrix $\left(
{\begin{array}{*{20}c}
   1 & 0 & 0  \\
   0 & 1 & 0  \\
   0 & 0 & 1  \\
\end{array}} \right)$.] separately to $\Gamma _1$, $\Gamma _2$, $\Gamma _3$ we have
three bases
\begin{equation}
\tilde \Gamma _1  = \{ I_{3 \times 3}^1 ,\lambda _1 ,\lambda _2
,\lambda _3 \}, \tilde \Gamma _2  = \{ I_{3 \times 3}^2 ,\lambda
_4 ,\lambda _5 ,\left( {\begin{array}{*{20}c}
   1 & 0 & 0  \\
   0 & 0 & 0  \\
   0 & 0 & { - 1}  \\
\end{array}} \right)\}, \tilde \Gamma _3  = \{ I_{3 \times 3}^3 ,\lambda _6 ,\lambda _7
,\left( {\begin{array}{*{20}c}
   0 & 0 & 0  \\
   0 & 1 & 0  \\
   0 & 0 & { - 1}  \\
\end{array}} \right)\},
\end{equation}
which can be mapped to the Lorentz group respectively as in the
last section. For example, the Lorentz group mapped from $\tilde
\Gamma _1$ has the form like Eq. (9)

\begin{equation}
\Lambda ^{\alpha \beta }  = \frac{1}{2}Tr(\lambda ^\alpha g\lambda
^\beta  g^ \dag),
\end{equation}
we note that now both the $\lambda ^\alpha$, $\lambda ^\beta$ and
the element $g$ of $SL$(2,C) group come from the set \{$I_{3
\times 3}^1 ,\lambda _1 ,\lambda _2 ,\lambda _3 $\}. Hereafter we
name the $\lambda ^\alpha$, $\lambda ^\beta$ in Eq. (15) \emph{the
basis of the mapping}. From Eq. (15), it can be proved that all of
the components $\Lambda ^{00}$, $\Lambda ^{0i}$, $\Lambda ^{i0}$
and $\Lambda ^{jk}$ are the same as those from Eq. (9). The
similar mappings from $\tilde \Gamma _2$ and $\tilde \Gamma _3$ to
Lorentz group will give the same result. We use these three bases
$\tilde \Gamma _1$, $\tilde \Gamma _2$ and $\tilde \Gamma _3$ to
signify the three families.

According to Eq. (15) and mimicking the matrices in Eq. (11), we
can construct the parallelism between the set \{$I_{3 \times 3}^1
,\lambda _1 ,\lambda _2 ,\lambda _3 $\} and Lorentz group as
follows
\begin{equation}
{ {\begin{array}{*{20}c}
   $Elements of SL (2, C)$  $~~~~~~~~~~~~~~~~~$& $Lorentz transformations$  \\
   \pm \left( {\begin{array}{*{20}c}
   {\cosh \frac{\kappa }{2}} & {\sinh \frac{\kappa }{2}} & 0  \\
   {\sinh \frac{\kappa }{2}} & {\cosh \frac{\kappa }{2}} & 0  \\
   0 & 0 & {\cosh \frac{\kappa }{2}}  \\
\end{array}} \right) $~~~~~~~~~~~~~~~~~$& \left( {\begin{array}{*{20}c}
   {\cosh \kappa } & {\sinh \kappa } & 0 & 0  \\
   {\sinh \kappa } & {\cosh \kappa } & 0 & 0  \\
   0 & 0 & 1 & 0  \\
   0 & 0 & 0 & 1  \\
\end{array}} \right) \\
 \pm \left( {\begin{array}{*{20}c}
   {\cosh \frac{\kappa }{2}} & {i\sinh \frac{\kappa }{2}} & 0  \\
   { - i\sinh \frac{\kappa }{2}} & {\cosh \frac{\kappa }{2}} & 0  \\
   0 & 0 & {\cosh \frac{\kappa }{2}}  \\
\end{array}} \right) $~~~~~~~~~~~~~~~~~$& \left( {\begin{array}{*{20}c}
   {\cosh \kappa } & 0 & { - \sinh \kappa } & 0  \\
   0 & 1 & 0 & 0  \\
   { - \sinh \kappa } & 0 & {\cosh \kappa } & 0  \\
   0 & 0 & 0 & 1  \\
\end{array}} \right) \\
   \pm \left( {\begin{array}{*{20}c}
   {Exp[\frac{\kappa }{2}]} & 0 & 0  \\
   0 & {Exp[ - \frac{\kappa }{2}]} & 0  \\
   0 & 0 & 1  \\
\end{array}} \right) $~~~~~~~~~~~~~~~~~$& \left( {\begin{array}{*{20}c}
   {\cosh \kappa } & 0 & 0 & {\sinh \kappa }  \\
   0 & 1 & 0 & 0  \\
   0 & 0 & 1 & 0  \\
   {\sinh \kappa } & 0 & 0 & {\cosh \kappa }  \\
\end{array}} \right) \\
\pm \left( {\begin{array}{*{20}c}
   {\cos \frac{\kappa }{2}} & {i\sin \frac{\kappa }{2}} & 0  \\
   {i\sin \frac{\kappa }{2}} & {\cos \frac{\kappa }{2}} & 0  \\
   0 & 0 & {\cos \frac{\kappa }{2}}  \\
\end{array}} \right) $~~~~~~~~~~~~~~~~~$& \left( {\begin{array}{*{20}c}
   1 & 0 & 0 & 0  \\
   0 & 1 & 0 & 0  \\
   0 & 0 & {\cos \kappa } & {\sin \kappa }  \\
   0 & 0 & { - \sin \kappa } & {\cos \kappa }  \\
\end{array}} \right) \\
   \pm \left( {\begin{array}{*{20}c}
   {\cos \frac{\kappa }{2}} & { - \sin \frac{\kappa }{2}} & 0  \\
   {\sin \frac{\kappa }{2}} & {\cos \frac{\kappa }{2}} & 0  \\
   0 & 0 & {\cos \frac{\kappa }{2}}  \\
\end{array}} \right)$~~~~~~~~~~~~~~~~~$& \left( {\begin{array}{*{20}c}
   1 & 0 & 0 & 0  \\
   0 & {\cos \kappa } & 0 & {\sin \kappa }  \\
   0 & 0 & 1 & 0  \\
   0 & { - \sin \kappa } & 0 & {\cos \kappa }  \\
\end{array}} \right) \\

 \pm \left( {\begin{array}{*{20}c}
   {Exp[i\frac{\kappa }{2}]} & 0 & 0  \\
   0 & {Exp[ - i\frac{\kappa }{2}]} & 0  \\
   0 & 0 & 1  \\
\end{array}} \right) $~~~~~~~~~~~~~~~~~$& \left( {\begin{array}{*{20}c}
   1 & 0 & 0 & 0  \\
   0 & {\cos \kappa } & {\sin \kappa } & 0  \\
   0 & { - \sin \kappa } & {\cos \kappa } & 0  \\
   0 & 0 & 0 & 1  \\
\end{array}} \right) \\

\end{array}} }
\end{equation}
where in the left column we have interpreted the \emph{generating
elements} of $SL$(2,C) with $3 \times 3$  matrices born from the
set \{$I_{3 \times 3}^1 ,\lambda _1 ,\lambda _2 ,\lambda _3$\},
the third diagonal elements are chosen so that when $\kappa  = 0$
they reduce to unit matrix. We note that in the second column,
despite the different signs in the second, the fourth, and the
sixth matrices, all of them have the reasonable forms like those
in Eq. (11), and meanwhile satisfy the condition $\eta _{\alpha
\beta } \Lambda ^{\alpha \mu } \Lambda ^{\beta \nu } = \eta ^{\mu
\nu }$, where $\eta ^{\alpha \beta }  = diag(1, - 1, - 1, - 1)$,
which is the basic requirement of Lorentz transformation.
Subsequently by making derivative to these matrices, one gets the
generators of Lorentz group.

The displayed matrices in (16) show up the mapping from $\tilde
\Gamma _1$ to Lorentz group. In nature, only one Lorentz group
should occur. So how the mappings from $\tilde \Gamma _2$ and
$\tilde \Gamma _3$ to Lorentz group manifest their existence is
worth studying. Merely in form, we denote the three equivalent
Lorentz groups, which are separately produced from $\tilde \Gamma
_1$, $\tilde \Gamma _2$ and $\tilde \Gamma _3$, by $L_1$, $L_2$,
$L_3$. Now we design a mapping that projects $\tilde \Gamma _2$
and $\tilde \Gamma _3$ also into $L_1$ in an extensive manner. The
mapping produces the degeneration of the groups as well as the
Lorentz violation. Based on Eq. (15), we choose the \emph{mapping
basis} $\lambda ^\alpha$, $\lambda ^\beta$ still from the set
\{$I_{3 \times 3}^1 ,\lambda _1 ,\lambda _2 ,\lambda _3$\}, but
the matrix $g$ from the combination of the elements in set $\tilde
\Gamma _2$ or $\tilde \Gamma _3$. For example, if the matrix $g$
comes from the linear combination of elements in $\tilde \Gamma
_2$, in analogy with the matrices in (16), the mapping result is
as follows
\begin{equation}
{ {\begin{array}{*{20}c}
   $Elements of SL (2, C)$  $~~~~~~~~~~~~~~~~~$& $Exended-Lorentz tansformations$  \\
  \pm\left( {\begin{array}{*{20}c}
   {\cosh \frac{\kappa }{2}} & 0 & {\sinh \frac{\kappa }{2}}  \\
   0 & {\cosh \frac{\kappa }{2}} & 0  \\
   {\sinh \frac{\kappa }{2}} & 0 & {\cosh \frac{\kappa }{2}}  \\
\end{array}} \right) $~~~~~~~~~~~~~~~~~$& \left( {\begin{array}{*{20}c}
   \chi  & 0 & 0 & 0  \\
   0 & \chi  & 0 & 0  \\
   0 & 0 & \chi  & 0  \\
   0 & 0 & 0 & \chi   \\
\end{array}} \right) \\
 \pm\left( {\begin{array}{*{20}c}
   {\cosh \frac{\kappa }{2}} & 0 & {i\sinh \frac{\kappa }{2}}  \\
   0 & {\cosh \frac{\kappa }{2}} & 0  \\
   { - i\sinh \frac{\kappa }{2}} & 0 & {\cosh \frac{\kappa }{2}}  \\
\end{array}} \right) $~~~~~~~~~~~~~~~~~$& \left( {\begin{array}{*{20}c}
   \chi  & 0 & 0 & 0  \\
   0 & \chi  & 0 & 0  \\
   0 & 0 & \chi  & 0  \\
   0 & 0 & 0 & \chi   \\
\end{array}} \right) \\
   \pm\left( {\begin{array}{*{20}c}
   {Exp[\frac{\kappa }{2}]} & 0 & 0  \\
   0 & 1 & 0  \\
   0 & 0 & {Exp[ - \frac{\kappa }{2}]}  \\
\end{array}} \right) $~~~~~~~~~~~~~~~~~$& \left( {\begin{array}{*{20}c}
   {\frac{1}{2}(1 + e^\kappa  )} & 0 & 0 & {\frac{1}{2}( - 1 + e^\kappa  )}  \\
   0 & {e^{\frac{\kappa }{2}} } & 0 & 0  \\
   0 & 0 & {e^{\frac{\kappa }{2}} } & 0  \\
   {\frac{1}{2}( - 1 + e^\kappa  )} & 0 & 0 & {\frac{1}{2}(1 + e^\kappa  )}  \\
\end{array}} \right) \\
\pm\left( {\begin{array}{*{20}c}
   {\cos \frac{\kappa }{2}} & 0 & {i\sin \frac{\kappa }{2}}  \\
   0 & {\cos \frac{\kappa }{2}} & 0  \\
   {i\sin \frac{\kappa }{2}} & 0 & {\cos \frac{\kappa }{2}}  \\
\end{array}} \right) $~~~~~~~~~~~~~~~~~$& \left( {\begin{array}{*{20}c}
   \chi  & 0 & 0 & 0  \\
   0 & \chi  & 0 & 0  \\
   0 & 0 & \chi  & 0  \\
   0 & 0 & 0 & \chi   \\
\end{array}} \right) \\
   \pm\left( {\begin{array}{*{20}c}
   {\cos \frac{\kappa }{2}} & 0 & { - \sin \frac{\kappa }{2}}  \\
   0 & {\cos \frac{\kappa }{2}} & 0  \\
   {\sin \frac{\kappa }{2}} & 0 & {\cos \frac{\kappa }{2}}  \\
\end{array}} \right) $~~~~~~~~~~~~~~~~~$& \left( {\begin{array}{*{20}c}
   \chi  & 0 & 0 & 0  \\
   0 & \chi  & 0 & 0  \\
   0 & 0 & \chi  & 0  \\
   0 & 0 & 0 & \chi   \\
\end{array}} \right) \\
\pm\left( {\begin{array}{*{20}c}
   {Exp[i\frac{\kappa }{2}]} & 0 & 0  \\
   0 & 1 & 0  \\
   0 & 0 & {Exp[ - i\frac{\kappa }{2}]}  \\
\end{array}} \right) $~~~~~~~~~~~~~~~~~$& \left( {\begin{array}{*{20}c}
   1 & 0 & 0 & 0  \\
   0 & {\cos \kappa } & {\sin \kappa } & 0  \\
   0 & { - \sin \kappa } & {\cos \kappa } & 0  \\
   0 & 0 & 0 & 1  \\
\end{array}} \right) \\

\end{array}} }
\end{equation}
where $\chi  = \cosh ^2 \frac{\kappa }{2}$.

Almost the same matrices would appear if in Eq. (15) the
\emph{mapping basis} still comes from $\tilde \Gamma _1$ but the
matrix $g$ comes from the linear combination of elements in
$\tilde \Gamma _3$, with some signs altered. In the column of
Exended-Lorentz transformations, we find that after performing
derivatives with respect to $\kappa$ and making $\kappa  = 0$,
only the third matrix leads to a nontrivial generator (the sixth
matrix produces a generator of original Lorentz group),

\begin{equation}
\left( {\begin{array}{*{20}c}
   {\frac{1}{2}} & 0 & 0 & {\frac{1}{2}}  \\
   0 & {\frac{1}{2}} & 0 & 0  \\
   0 & 0 & {\frac{1}{2}} & 0  \\
   {\frac{1}{2}} & 0 & 0 & {\frac{1}{2}}  \\
\end{array}} \right).
\tag{18a}
\end{equation}
We call the matrix an extension of Lorentz generators, remembering
that $\left( {\begin{array}{*{20}c}
   0 & 0 & 0 & 1  \\
   0 & 0 & 0 & 0  \\
   0 & 0 & 0 & 0  \\
   1 & 0 & 0 & 0  \\
\end{array}} \right)$ is also a generator of original Lorentz group, its
effective part can be written as
\begin{equation}
\left( {\begin{array}{*{20}c}
   1 & 0 & 0 & 0  \\
   0 & 1 & 0 & 0  \\
   0 & 0 & 1 & 0  \\
   0 & 0 & 0 & 1  \\
\end{array}} \right).
\tag{18b}
\end{equation}
The Exended-Lorentz matrices due to the mapping from $\tilde
\Gamma _3$ give the same conclusion. In Eq. (15), if $\lambda
^\alpha$, $\lambda ^\beta$ is chosen from the set $\tilde \Gamma
_2$ ($\tilde \Gamma _3$), and the matrix $g$ from the combination
of the elements in set $\tilde \Gamma _1$ or $\tilde \Gamma _3$
($\tilde \Gamma _1$ or $\tilde \Gamma _2$), we will gain the same
results. Without losing generality, in what follows we confine
ourselves to developing the results from Eqs. (17), (18). From
equation (18b) we can see the structure of extended Lorentz group:
the obtained remnant generator commutes with all of the other
generators in Lorentz group, the Closure condition is obviously
satisfied. We give more discussion on the group structure at the
end of next section.
\section{The Extension of Kinetic Vertex of Dirac Equation}
Free leptons in each of families should satisfy the Dirac
equation,
\begin{equation}
\gamma _\mu  i\partial ^\mu  \psi  = m\psi, \tag{19}
\end{equation}
but this equation will not be accurate if the above extension of
Lorentz generators is acceptable. We will elucidate the problem
and resolve it in this section.

We here define the $\gamma _\mu$ as \emph{kinetic vertex} of the
Dirac equation. Performing the Lorentz transformation in Eq. (7)
on both sides of Dirac equation and at the same time assuming that
$\psi$ transforms according to
\begin{equation}
\psi (x') = S^{ - 1} \psi (x), \tag{20}
\end{equation}
where $S$ is a nonsingular $4\times 4$ matrix, then one concludes
[17]
\begin{equation}
S^{ - 1} \gamma _\mu  S = \gamma _\nu  \Lambda _{\;\,\mu} ^\nu.
\tag{21}
\end{equation}
Let's introduce the explicit form $\Lambda _{\;\,\mu} ^\nu   =
\delta _{\;\,\mu }^\nu   + \omega _{\;\,\mu }^\nu$, where $\omega
_{\;\,\mu }^\nu$ comes from the generators of Lorentz group and
has the expression $\omega _{\;\,\mu }^\nu   = g^{\nu \lambda }
\varepsilon _{\lambda \mu }$[17], and $\varepsilon _{\lambda \mu
}$ is an infinitesimal antisymmetric tensor. Substituting this
explicit form into Eq. (21) and making some changes, we have
\begin{equation}
\gamma _\mu  S - S\gamma _\mu   = [\gamma _\mu  ,S] = S\omega
_{\;\,\mu }^\nu  \gamma _\nu. \tag{22}
\end{equation}
Furthermore if we take into account the infinitesimal parameter
for the transformation $S$
\begin{equation}
S = 1 + \varepsilon _{\mu \nu } S^{\mu \nu }, \tag{23}
\end{equation}
then to first order, Eq. (22) can be written
\begin{equation}
[\gamma _\mu  ,\varepsilon _{\rho \sigma } S^{\rho \sigma } ] =
\omega _{\;\,\mu }^\nu  \gamma _\nu, \tag{24}
\end{equation}
accordingly we find the solution of Eq. (24) must be
\begin{equation}
S^{\mu \nu }  = \frac{1}{2}\gamma ^\mu  \gamma ^\nu. \tag{25}
\end{equation}
The above formalism in this section are based on the conventional
Lorentz generators. If a generator like matrix (18b) appears, then
to reserve the Dirac equation form
\begin{equation}
\Pi _\mu  \partial ^\mu  \psi  = m\psi, \tag{26}
\end{equation}
in which $\Pi _\mu$ is a general form of kinetic vertex being the
linear combination of $\gamma$ matrices, its consequence Eq. (24)
must be altered correspondingly. Roughly we can rewrite the
general form of Eq. (24) analog as
\begin{equation}
[\Pi _\mu  ,\bar \varepsilon _{\rho \sigma } S^{\rho \sigma } ] =
\bar \omega _{\;\,\mu }^\nu  \Pi _\nu, \tag{27}
\end{equation}
where $\bar \omega _{\;\,\mu }^\nu$ now is unit matrix, and $\bar
\varepsilon _{\mu \nu }$ may not be antisymmetric any longer. Then
it is found when
\begin{equation}
\Pi _\mu   = \gamma _\mu  (1 + \gamma _5 ), \bar \varepsilon _{\mu
\nu } S^{\mu \nu }  = \frac{{\bar \omega _{\;\,\mu }^\nu
}}{2}\gamma _5, \tag{28}
\end{equation}
the equation (27) can be satisfied. This means that the
\emph{kinetic vertex} of Dirac equation has been extended to
include $\gamma _\mu  (1 + \gamma _5 )$, and the Lorentz
generators in spinor space \{$\frac{1}{2}\gamma ^\mu  \gamma
^\nu$\}
 have been extended to include $\gamma _5$. So
now the Dirac equation can be written
\begin{equation}
\gamma _\mu  [(1 + w) + w\gamma _5 ]i\partial ^\mu  \psi  = m\psi,
\tag{29}
\end{equation}
where $w$ is a small parameter determined according to specific
situation. If in Eq. (29) $\det [(1 + w) + w\gamma _5 ] \ne 0$
 and thus $[(1 + w) + w\gamma _5 ]$ has its inverse matrix, the changes appears to kinetic vertex
can be transferred to mass term. In that context by making $w$
carry index like $w_F^{\mu \nu }$ or $w_F^{\mu}$($F$ stands for
flavor indices to make summation), the discussions in the Ref. [9]
can be followed to get the neutrino oscillations as well as mass
differences.

We call the set \{$\gamma _5$, \{$\frac{1}{2}\gamma ^\mu  \gamma
^\nu$\}\} an extended Lorentz group (in spinor representation):
the group Closure condition is kept by recognizing that the
product of $\gamma _5$ and any $\frac{1}{2}\gamma ^\mu  \gamma
^\nu$ is still in the set \{$\frac{1}{2}\gamma ^\mu  \gamma
^\nu$\}, and their commutator $[\gamma _5,\frac 12\gamma _\mu
\gamma _\nu ]=0$; $\gamma _5$ and \{$\frac{1}{2}\gamma ^\mu \gamma
^\nu$\} actually form a group, in which $\gamma _5$ turns out to
be an identity element. And the anti-symmetry of generators'
matrix elements of Lorentz group is lost due to the entering of
$\gamma _5$.

\section {Summary and Discussion}

In this paper we propose a mapping on the basis of Eq. (15) to
elucidate the mutual impacts of three families. The families are
thought to have $SL$(3, C) [which can be graded by $SU(3)_F$]
symmetry but show their effect by Lorentz-like invariance. The
group $SU(3)_F$ is divided into three $SU$(2) subgroups, each of
which can be mapped into a proper Lorentz group, resembling the
way we treating the mapping between group $SL$(2, C) and Lorentz
group. In this way three families are mapped to three equivalent
Lorentz groups (\emph{Lorentz zones}). When the mapping from group
$SL$(3, C) to Lorentz group is constructed by choosing one $SL$(2,
C) subgroup as \emph{mapping basis}, the other two subgroups
display their impacts only by one more generator additional to
that of the original Lorentz group.

If we extend the Dirac equation (19) to include interaction, then
it is written
\begin{equation}
\gamma _\mu  (i\partial ^\mu   - gA^\mu  )\psi  = m\psi. \tag{30}
\end{equation}
The extension of kinetic vertex in Eq. (29) is obviously
applicable to the interaction term $g\gamma _\mu  A^\mu$ too,
\begin{equation}
\gamma _\mu  A^\mu   \to \gamma _\mu  [(1 + w) + w\gamma _5
]A^\mu. \tag{31}
\end{equation}
The Eq. (31) provides a way the axial-vector interaction $\gamma
_\mu  \gamma _5 A^\mu$ arises. Reversing the causality, one may
form the idea that the appearance of axial-vector form in weak
interaction determines the triplicity of lepton families.

At the unification scale, the families' symmetry should be broken
to produce mass differences and mass hierarchy of leptons and
quarks [8]. The Lorentz invariance might also be broken at this
scale to accommodate the inclusion of gravity in a unified theory
[9]. So far, by virtue of our mapping, we guess that the two
symmetries are possibly dependent on each other at broken scale,
as well as dependent on the weak interaction, as argued above.
Finally, we stress that the proposed Lorentz extension (Lorentz
violation) here should be detected (confirmed or denied) under
interactions other than Electrodynamics.

\section*{Acknowledgments}

This work is supported by National Natural Science Foundation of
China under Granted No.10675054 and No.10775059.


\begin{thebibliography}{99}


\bibitem{1} A. B. Bracic and N. S. M. Borstnik, Phys. Rev. D 74, 073013
(2006).
\bibitem{2} N. M. Borstnik, H. B. Nielsen, J. Math. Phys.
44, 4817 (2003).
\bibitem{3} Y. Kawamura, T. Kinami, K. Oda, Phys. Rev. D 76, 035001 (2007), also at arXiv:hep-ph/0703195.
\bibitem{4} T. Watari and T.
Yanagida, Phys. Lett. B, 532, 252 (2002).

\bibitem{5} Bob Holdom,
JHEP 08 (2006) 076.

\bibitem{6} M. Cveti\v{c}, G. Shiu, and A. M. Uranga, Phys. Rev. Lett. 87,
201801 (2001).
\bibitem{7}Ernest Ma, Mod. Phys. Lett. A 20, 1953 (2005).
\bibitem{8} Ernest Ma, Non-Abelian Discrete Flavor Symmet,
arXiv:0705.0327 [hep-ph]; and relevant references therein.
\bibitem{9} V. A. Kostelecky¡ä and M. Mewes, Phys. Rev. D 69, 016005 (2004).
\bibitem{10}Ashok Das, J. Gamboa, J. L¨®pez-Sarri¨®n, and F. A. Schaposnik,
Phys. Rev. D 72, 107702 (2005).
\bibitem{11} A. G. Cohen and S. L.
Glashow, Phys. Rev. Lett. 97, 021601 (2006).

\bibitem{12} Peter O. Hess, Walter Greiner, arXiv:0705.1233 [hep-th].
\bibitem{13}Moshe Carmeli and Shimon Malin, Theory of Spinors: An
Introduction, World Scientific Publishing Co. Pte. Ltd. (2000).
\bibitem{14} H de Guise, J. Patera, R. T. Sharp, J. Math. Phys. 41 (2000)
4860.

\bibitem{15} Havlicek M, Patera J, Pelantova E, Tolar J, JOURNAL OF
NONLINEAR MATHEMATICAL PHYSICS 11: 37-42 Suppl. S 2004.

\bibitem{16} J. Hrivnak, P. Novotny, J. Patera, J. Tolar, Linear Algebra
and its Applications, 418, (2006) 498, also:
arXiv:math-ph/0509033.
\bibitem{17} F. Mandl, G. Shaw, Quantum Field
Theory (Jhon Wiley and Sons Ltd., 1984), Appendix A.7.

\end{thebibliography}
\end{document}